\documentstyle[prb,aps,psfig]{revtex}
\begin{document}
\draft
%%%%%%%%%%%%%%%%%%
\twocolumn[\hsize\textwidth\columnwidth\hsize\csname
@twocolumnfalse\endcsname
%%%%%%%%%%%%%%%%%%
\widetext
\title{\bf{Non linear sigma models and quantum spin systems}}
\author{
Antonio S. Gliozzi$^{1,2}$
and Alberto Parola$^{1,3}$
}
\address{
$^1$ Istituto Nazionale per la Fisica della Materia, Como \\
$^{2}$ Dipartimento di Fisica, Universit\'a di Milano, Via Celoria 16,
Milano, Italy \\
$^3$ Dipartimento di Scienze Fisiche, Universit\'a dell'Insubria, Via Valleggio 11
Como, Italy }
\maketitle
\begin{abstract}
Microscopic models of quantum antiferromagnets are investigated on the basis of
a mapping onto effective low energy hamiltonians. Lattice effects are carefully
taken into account and their role is discussed. We show that the presence of
an external staggered magnetic field modifies in a non trivial 
way the usual mapping onto 
the non linear sigma model, leading to the appearance of new terms, 
neglected in previous works.
Our analysis is compared with Lanczos diagonalizations
of $S=1$ Heisenberg chains in a staggered field, confirming the
validity of the single mode approximation for the evaluation of 
the dynamical structure factor.  The results are relevant for 
the interpretation of experiments in quasi-one dimensional compounds. 
Microscopic realizations of SU(4) spin chains are also discussed in the
framework of spin-orbital lattice systems.
The low energy physics is shown to be described by 
sigma models with topological angle $\theta$ in one dimension. 
This mapping strongly suggests that the one dimensional 
CP$^3$ model (with $\theta=\pi$) undergoes a second order phase transition
as a function of the coupling.
\end{abstract}
\pacs{75.10.Jm 05.70.Jk 75.40.Cx}
]
\narrowtext
\section{Introduction}
Quantum spin models in low dimensionality are currently used to 
describe the magnetic properties of several materials including
rare earths, organic compounds, transition metals and copper oxides.
Available experimental data in these systems, especially
magnetic resonance and neutron scattering, allow to establish 
the local structure as well as the most relevant features of the
long range correlations. 
These studies generally show that magnetic materials can
be accurately described by short range spin hamiltonians,
at least in temperature regimes where the effects of disorder, 
the presence of spatial anisotropies or of dipolar interactions
are negligible. The possibility to use simple spin hamiltonians
to understand the physics of real systems has always been 
the drawing force for the development of more and more accurate 
methods to study the phase diagram of these models.
Purely analytical techniques, like spin-waves expansions or mean
field theories \cite{swt} have shown to be quite accurate when
magnetic ordering is present. Numerical methods, based on 
series expansions \cite{series}, Quantum Monte Carlo simulations
\cite{simul} or Density Matrix Renormalization Group (DMRG)\cite{white}
have been successfully applied also to finite temperatures, to frustrated 
models and to quasi one dimensional systems, where fluctuations 
play a key role in determining the physics of the model.
A complimentary class of theoretical methods, which has been developed 
for investigating the effects of quantum and thermal fluctuations 
in many body systems, is known as the semiclassical approach and
includes different techniques which have been used
to provide a physical interpretation to experimental 
and simulation data \cite{chn,tognetti,gianinetti}.

Semiclassical theories have been widely used in framework of 
quantum statistical mechanics since a long time. The mapping of quantum models,
in particular spin systems, onto classical effective statistical
models has been extremely useful in understanding many relevant
features of phase diagrams and the possible occurrence of quantum 
critical points. For instance, the one dimensional Ising model in a 
transverse magnetic field
maps to the (anisotropic) two dimensional Ising model: its exact 
solution therefore provides a simple way to investigate the 
critical properties of the Ising universality class \cite{sachdev}. 
The quantum to classical mapping can be usually justified on microscopic grounds
only in the low energy and long wavelength limit, where the short
range features of the original quantum model become irrelevant.
Therefore, strictly speaking, the use of semiclassical methods in
quantum statistical mechanics is restricted to regions characterized
by a diverging correlation length and gapless low energy excitations. 
However, the microscopic quantum hamiltonian is 
often assumed to be faithfully represented by its classical counterpart
in a large portion of the phase diagram provided the coupling constants
present in the classical lagrangian are suitably {\it renormalized} due
to quantum (short range) fluctuations \cite{sachdev}.
This expectation has been beautifully confirmed by the study of 
Heisenberg models in low dimensionality. In particular, exact 
solutions and numerical analysis of one dimensional antiferromagnets found
exponentially decaying correlations and gapped excitation 
spectrum in integer spin chains. Conversely, 
semi-integer spin chains turned out to be gapless with power-law correlations.
This picture fully agrees with the conjecture by Haldane, based
on the presence of a topological term in the semiclassical action
of one dimensional spin chains \cite{haldane}. Moreover, quantitative 
analysis of the three dimensional Non Linear Sigma Model (NL$\sigma$M)
found a phase diagram which compares favorably with experiments on
two dimensional antiferromagnets \cite{chn,beard,troyer} suggesting 
that semiclassical approaches can be directly used for the interpretation
of experimental data. 

The semiclassical mapping for Heisenberg antiferromagnets
has been also employed to study the effects of an external magnetic field 
on the model. Of particular physical relevance is the role of a 
staggered field, which directly couples to the order parameter:
Such a staggered field can be realized in certain quasi-one dimensional 
spin one compounds where, by lowering the temperature, 
rare earths magnetic ions undergo a N\'eel transition to a three 
dimensional antiferromagnet which generates an alternating
magnetic field acting on the Ni$^{2+}$ chains
\cite{maslov}. A similar mechanism has been also invoked for
the interpretation of quasi one dimensional spin one-half chains
\cite{oshikawa}. However, the effective classical action
used in the literature has not been explicitly derived from the 
quantum hamiltonian and we believe it should be reconsidered.
In order to extract quantitative information from this NL$\sigma$M
a ``Single Mode Approximation" (SMA) is usually adopted \cite{maslov}
but a detailed numerical study of its accuracy in this case is still missing,
even if analytical studies \cite{morandi} suggest that it may be 
justified only for the transverse channel.
In this paper, we present a microscopic derivation of the semiclassical action.
We obtain an effective low energy theory different from the one commonly adopted
in the literature. This theory is then analyzed in the weak and
strong field limit and the results are compared to Lanczos diagonalizations.
 
Other magnetic materials where the use of semiclassical methods may be 
suggestive are spin-orbital systems, like C$_{60}$ compounds \cite{auerbach}
or transitional metal oxides \cite{pati}, in which 
orbital degeneracy is present. 
The special points where the model has an enlarged SU(4) symmetry
are particularly important in order to understand the phase diagram 
of the model \cite{santoro,li,itoi} and deserve a detailed analysis.
We therefore apply the semiclassical mapping to the two microscopic models 
presenting this symmetry. We show that in one case the model maps 
straightforwardly to the CP$^3$ NL$\sigma$M (with topological term in the one
dimensional limit) while in the other case, we have been able to carry
out the mapping only in the special case of one dimension, where again 
we find the same semiclassical action at a different effective coupling.
Interestingly, both lattice hamiltonians have been exactly solved
in D=1 \cite{martins,sutherland} with very different results: the
first model has broken symmetry and gapped excitations, while in
the second case it is gapless and critical. We believe that this mapping 
provides an important clue to the understanding 
the phase diagram of CP$^n$ models,
which are shown to undergo a phase transition as a function of the coupling
constant.

\section{Semiclassical approach}
In this Section we briefly review and generalize a method, 
proposed few years ago \cite{ap}, for the
derivation of low energy effective actions in bipartite spin systems. The main advantage
of this technique, with respect to the original 
procedure developed by Haldane \cite{haldane},
is the possibility to keep track of lattice effects and to make direct contact with other
useful microscopic approaches, like spin wave theory. The method will be later applied to
Heisenberg models in an external field and to spin-orbital systems.

Be $H$ the hamiltonian of a spin model on a bipartite lattice.
For instance, the celebrated Heisenberg antiferromagnet is described by
\begin{equation}
{\cal H} =\sum_{R\in B}\sum_{\delta} {\bf S}_R \cdot  {\bf S}_{R+\delta}
\label{ham1}
\end{equation}
where ${\bf S}$ are spin operatrs,
the site index $R$ runs over the sublattice labeled by $B$
and $\delta$ 
is a primitive vector on the lattice connecting nearest neighbor sites.
In order to evaluate the partition function $Z={\rm Tr} \exp(-\beta {\cal H})$ we adopt the
usual coherent states formalism: we first split the
interval $(0,\beta)$ in a large number $N$ of Trotter (time) slices, then insert 
at each imaginary time $t$ a resolution of unity based on 
coherent states defined at each lattice site $R$. Here 
we follow the standard $O(3)$ notation $|{\bf\Omega}(R,t)>$:
In spin $S$ models the coherent states are labeled by the unit vector 
${\bf\Omega}(R,t)$ and are characterized by the requirement that 
$<{\bf\Omega}|\,{\bf S}\,|{\bf\Omega}>=S {\bf\Omega}$. These states may be explicitly 
obtained by a suitable rotation in spin space of the highest eigenvector of the $S_z$
operator. The partition function can be therefore written as:
\begin{eqnarray}
Z&=&\int {\cal D} {\bf \Omega}(R,t) \,e^{-S_{eff}} \nonumber \\
S_{eff}&=&\int_0^\beta dt \Big [
S^2\sum_{R\in B} \sum_{\delta} {\bf \Omega}(R,t) \cdot  {\bf \Omega}(R+\delta,t) -
\nonumber \\
&&\sum_R <{\bf\Omega}(R,t)|\dot{\bf\Omega}(R,t)> \Big ]
\label{action}
\end{eqnarray}
Up to this point, the underlying lattice structure is fully present 
in the functional form (\ref{action}) and no approximation has been 
introduced. However, the imaginary Wess-Zumino term 
\begin{equation}
<{\bf\Omega}|\dot{\bf\Omega}>=iS\,{({\bf\Omega}\times\dot{\bf\Omega})_z\over
1+\Omega_z}
\end{equation}
prevents a simple classical interpretation of the effective action. 
In order to obtain a mapping onto a physically transparent classical statistical model, 
it is convenient to exploit the bipartite nature of the lattice
by explicitly tracing out the degrees of freedom defined on 
sublattice $B$.
This procedure can be performed analytically because the variables we are 
integrating out are coupled only to the classical external fields
${\bf\Omega}(R,t)$ defined on the other sublattice (sublattice $A$). 
Therefore this step just 
requires the solution of the {\it single site} time dependent problem:
\begin{eqnarray}
e^{-F[{\bf B}(R,t)]}={\rm Tr} \,U_\beta \nonumber\\
{dU_t\over dt} = - {\bf K}(R,t) \,U_t \nonumber \\
{\bf K}(R,t)={\bf S}\cdot{\bf B}(R,t)  \nonumber \\
{\bf B}(R,t)=S \sum_\delta {\bf\Omega}(R+\delta,t)
\label{onsite}
\end{eqnarray}   
In terms of the local free energy functional $F[{\bf B}(R,t)]$, the effective action
becomes:
\begin{equation}
S_{eff}= \sum_{R\in B} F[{\bf B}(R,t)] -
\int_0^\beta dt\sum_{R\in A} <{\bf\Omega}(R,t)|\dot{\bf\Omega}(R,t)> 
\label{act}
\end{equation}
Notice that now the effective action just depends on the field ${\bf\Omega}(R,t)$ on 
sublattice $A$.
The solution of the problem defined in Eq. (\ref{onsite}) can be obtained 
within perturbation theory
for {\it slowly varying} field configurations. This requirement is 
satisfied in the (semiclassical) large $S$ limit and represents the 
only, {\it low energy} approximation we need to introduce 
in the evaluation of the effective action. We notice that the lattice structure 
is not involved in this semiclassical analysis and the {\it long wavelength}
approximation is actually unnecessary within our method. 
Explicitly, we obtain the following free energy functional:
\begin{eqnarray}
F[{\bf B}(t)]&=& \int_0^\beta dt \left\{\epsilon_0(t) + \Gamma_{00}(t) - 
\int_0^t dt^\prime
\Gamma_{01}(t)\Gamma_{10}(t^\prime)\right \} \nonumber\\
\Gamma_{ij}(t)&=&<u_i(t)|\dot u_j(t)>\exp\left \{ \int_0^t 
dt^\prime \left [ \epsilon_i(t^\prime)-\epsilon_j(t^\prime)\right ] \right \}
\label{free}
\end{eqnarray}
where $\epsilon_i(t)$ is the $i^{th}$ instantaneous eigenvalue of ${\bf K}(R,t)$
and $|u_i(t)>$ is the corresponding eigenstate. In order to obtain 
this expression we have assumed that the ground state of 
${\bf K}(R,t)$ $|u_0(t)>$ is
non degenerate at every time $t$. The terms shown in Eq. (\ref{free}) are correct up
to second order in time derivatives, as usual in semiclassical approximation.
The long wavelength limit of the Berry phase term $\Gamma_{00}$ in $F[{\bf B}(t)]$ 
has been shown in Ref. \cite{ap}
to exactly compensate the Wess-Zumino contribution in the effective action
(\ref{act}). As a result, the low energy {\it and} long wavelength effective action, 
becomes a non-linear sigma model with,
possibly, a topological $\theta$ term coming from the residual space 
dependence of the Berry phase $\Gamma_{00}$.
 
Now we are in the position to apply this method to specific lattice hamiltonians:
The procedure we have just outlined requires the explicit solution of the 
instantaneous eigenvalue problem
defined by the on site hamiltonian ${\bf K}(R,t)$ (\ref{onsite}), the evaluation of
the terms $\Gamma_{ij}(t)$ in Eq. (\ref{free}) and finally the substitution of 
the results into the form of the action (\ref{act}).

\section{Heisenberg model in an external field}

It is now established that the ground state properties
and the low lying excitation spectrum of antiferromagnetic 
Heisenberg chains strongly depend on the value of the on site magnetic moment. 
Semi-integer spin chains have power law correlations and
gapless spectrum while, for integer spin, correlations 
decay exponentially and a gap to all excitations is present \cite{review}
($\Delta=0.41048$ for $S=1$).
This different behavior was first found by Haldane\cite{haldane} 
via a semi-classical mapping onto the $O(3)$ non-linear $\sigma$-model. 
This conjecture was later confirmed by numerical studies based on 
Lanczos diagonalizations \cite{parkinson}, DMRG \cite{white} and 
MC simulations \cite{night}. 

Remarkably, these idealized models have also an experimental counterpart: neutron 
scattering experiment on quasi-one dimensional materials such as 
{\it ${\rm Ni(C_2H_8N_2)_2NO_2(ClO_4)}$} (NENP), 
confirmed\cite{neutron} that the essential physics of these systems 
is well described by a 
simple $S=1$ Heisenberg hamiltonian that couples neighboring spins 
antiferromagnetically and takes account 
of the single-ion anisotropy by including an on site term $D\,(S_i^z)^2$. 
The predicted Haldane gap has been measured with great accuracy in these experiments
showing good agreement with theoretical and numerical\cite{sorensen,lanczos,mc} 
predictions.

The effects of a magnetic external perturbation on an antiferromagnetic
chain, frequently gives rise to unexpected interesting phenomena. This is the
case of Cu benzoate, a quasi one dimensional
$S=1/2$ antiferromagnet displaying a gapped 
excitation spectrum in an applied uniform field.
Oshikawa and Affleck \cite{oshikawa} interpreted the experimental 
findings on the basis of an Heisenberg hamiltonian 
where spins are coupled to a weak effective staggered field.
This microscopic model gives a field dependence for the gap 
which agrees with experimental data.

More recently, interest has been focussed on the study of the effects of external 
fields on $S=1$ systems. In the case of a uniform magnetic field, the lowest triplet 
excitation states are split into a transverse and a longitudinal mode and the gap 
closes at a critical field $H_c$, 
where Bose condensation of magnons takes place\cite{s1}. The synthesis of 
compounds of the form ${\rm R_2BaNiO_5}$ (where R stands for a magnetic rare earth) 
allowed to study
the effects of a {\it staggered} magnetic field on the quasi-one dimensional 
chain of spin one Ni$^{2+}$ ions. 
The magnetic moment of the Ni$^{2+}$ couples with the R$^{3+}$ ions that are 
ordered antiferromagnetically below a N\'eel temperature $T_N$ (typically $16 K\lesssim T_N \lesssim 80 K$).
This three dimensional antiferromagnetic matrix generates an effective staggered magnetic field on the 
Ni$^{2+}$ chains whose intensity can be tuned by
varying the temperature below $T_N$. 
In this way, experiments have been able to investigate the effects 
of a staggered field on  the Haldane gap, the staggered magnetization and 
the susceptibility \cite{zhel,ray}.

Stimulated by these experiments, few analytical and numerical studies attempted a theoretical
analysis of spin chains in staggered fields by use of semiclassical mappings 
\cite{maslov,morandi} and DMRG \cite{yulu}.
While qualitative agreement
can be easily attained, some discrepancy still remains between the NL$\sigma$M approach
and DMRG findings, noticeably on the form of correlation functions. 
This circumstance is rather surprising in light of the very nice agreement between 
the NL$\sigma$M predictions and numerical data for the pure Heisenberg chain \cite{sorensen}.

The strong field limit is particularly simple because, for every $S$,
the ground state can be accurately described by a N\'eel state 
with gaussian transverse fluctuations. For such a problem, 
spin wave theory (SWT) can be applied also in one dimension
giving a transverse spectrum of the form:
\begin{equation}
\epsilon_k=S\,\sqrt{(|{\bf H}|/S+2D)^2-4\gamma_k^2}
\label{swtd}
\end{equation}
where ${\bf H}$ is the staggered field, $D$ is the space dimensionality 
and $\gamma_k=\sum_{i=1}^D \cos k_i$.
To leading order, the transverse dynamical correlations in imaginary time
predicted by SWT have single mode character:
\begin{equation}
S_{\perp}(k,\omega)=S^2{|{\bf H}|/S+2(D -\gamma_k) \over \omega^2+\epsilon_k^2}
\end{equation}
The longitudinal correlations may be expressed as convolutions of the transverse ones,
implying that no longitudinal branch of elementary excitations is present. 
According to the SWT approach, in the strong field limit, the longitudinal gap 
saturates at twice the transverse one.

Much more subtle is the weak field case, where quantum fluctuations strongly 
contrast the onset of a magnetically ordered state.
In order to understand this limit, we re-examine the derivation of the NL$\sigma$M
for a spin $S$ chain in an external staggered field on the 
basis of the method sketched in Section II.

The microscopic hamiltonian we consider here is just the 
antiferromagnetic Heisenberg model in an external
field ${\bf H}$ that can be taken either uniform or staggered:
\begin{equation}
{\cal H} =\sum_{R \in B} {\bf S}_R \cdot  \big[\sum_\delta{\bf S}_{R+\delta}+
{\bf H}\big]\pm{\bf H}\cdot\sum_{R\in A}{\bf S}_R 
\label{hamiltonian}
\end{equation}
where the conventions are the same as in (\ref{ham1}), and upper (lower) sign refers to uniform 
(staggered) applied field. 
Following the derivation of the previous Section,
we factorize the problem in the two sublattices ($A$ and $B$) and write the effective action as:
\begin{eqnarray}
S_{eff}=-\int_0^\beta dt &&\sum_{R\in A} \bigg\{<{\bf\Omega}(R,t)|
\dot{\bf\Omega}(R,t)>
\mp S{\bf H}\cdot{\bf\Omega}(R,t)\bigg\}  \nonumber\\
&&+\sum_{R\in B} F[{\bf B}(R,t)]
\label{partstag} 
\end{eqnarray}
where now
\begin{equation}
{\bf B}(R,t)=S\sum_\delta{\bf\Omega} (R+\delta,t)+{\bf H}.
\label{field}
\end{equation}
Therefore, we formally reduce to the same 
one-body problem as stated in Eq. (\ref{onsite}). 
This can be solved perturbatively in the case 
of zero temperature ($\beta\to\infty$), and slowly varying effective fields ${\bf B}$. The
resulting functional $F[{\bf B}]$ is then correct up to  
second order in space and time derivatives. 
In this context we stress that this perturbative treatment is justified only in the low-energy 
limit which can be physically accessed in gapless systems. 
In one dimension such a requirement is 
satisfied in half-integer spin chains while, for integer spins, 
it holds only in the large $S$ limit.
In fact, a perturbative renormalization group analysis, in zero external field, 
predicts the exponential dependence \cite{polyakov} $\Delta\sim \exp{-(\pi S)}$.
This shows that the existence of the gap is a purely quantum effect: it vanishes
on approaching the  
{\it classical} limit ($S\to\infty$) where it is still justified to 
derive the effective action perturbatively also for integer spin systems.
Interestingly, in the {\sl strong} field limit, the one body problem can be
again easily solved by considering small oscillations of the vector ${\bf \Omega}$
about the direction of the magnetic field. The resulting effective action, to quadratic
order in the amplitude of the oscillations, exactly reproduces all the 
lowest order results of SWT, including lattice effects. 

Specializing Eq.(\ref{free}) to the form of the effective field (\ref{field}) we obtain
\begin{equation}
F[{\bf B}(R,t)]=\int_0^\beta dt\left\{
S\frac{ |{\bf \dot m}(R,t)|^2}{2\,|{\bf B}(R,t)|}-S|{\bf B}(R,t)|+
\Gamma_{00}[{\bf m}]\right\}
\label{sefstag}
\end{equation}
where ${\bf m}=\frac{ {\bf B}}{|{\bf B}|}$. 
The first term comes from the time integration 
of Eq.(\ref{free}) and the second one is the 
ground state eigenvalue $\epsilon_0(t)$.  
Now we perform the continuum limit of the expression
(\ref{sefstag}) assuming that the relevant configurations ${\bf\Omega}(R,t)$ are 
slowly varying functions of space on the scale set by the lattice spacing $a$.
To lowest order in spatial fluctuations we have
${\bf B}(R,t)=S\,q\,{\bf\Omega} (R_0,t)+{\bf H}$ where $R_0=R-\hat x$ is a reference site 
belonging to sublattice $A$ ($\hat x$ is the primitive vector pointing
in the $x$ direction)
and $q=2D$ is the number of nearest neighbors of a hypercubic lattice
in dimension $D$. In the weak field limit, we just need to keep terms up to second order
in  $\delta {\bf B}(R,t)={\bf H}+S\sum_\delta[{\bf\Omega}(R+\delta,t)-{\bf\Omega}(R_0,t)]$.
By expanding ${\bf m}(R,t)$ we obtain:
\begin{eqnarray}
\delta{\bf m}&(&R,t)={\bf m}(R,t)-{\bf\Omega}(R_0,t)\nonumber\\
&&=\frac{\delta {\bf B}(R,t)}{S\,q}-
\frac{{\bf\Omega}(R_0,t)\cdot\delta {\bf B}(R,t)}{S\,q} +O(\delta B^2)
\end{eqnarray}
This leads to an approximation of the Berry phase that, to lowest order
cancels the Wess-Zumino term in the effective action leaving a residual contribution:
\begin{eqnarray}
\Gamma_{00}[{\bf m}] &-& \Gamma_{00}[{\bf \Omega}(R_0,t)] \nonumber\\
&\simeq&iS\delta {\bf m}(R,t)\cdot\big({\bf m}(R,t)\times{\bf\dot m}(R,t)
\big)  \nonumber\\
&\simeq&\frac{i}{q} {\bf H}\cdot\big({\bf \Omega}(R_0,t)
\times{\bf \dot\Omega}(R_0,t)\big)
\label{gammaoo}
\end{eqnarray}
In one dimension, the usual topological term, arising from the spatial 
derivative of ${\bf\Omega}$
present in $\delta{\bf B}$, also appears besides the contributions 
shown in Eq. (\ref{gammaoo}). 
By taking the continuum limit, 
we finally get the form of an effective action in which one half of the 
degree of freedom have been integrated out:
\begin{eqnarray}
&S_{eff}&=\int dt \int {dR\over 2a^D}
\bigg [\frac{1}{2q}\left(\dot{\bf \Omega}(R,t)+
i\,{\bf H}\times{\bf\Omega}(R,t)\right )^2 + \nonumber\\
&& a^2S^2 |\nabla_R {\bf \Omega}|^2
-(1 \mp 1) S{\bf H}\cdot{\bf \Omega}(R,t)\bigg ] +2\pi i S \,Q
\label{acteffstag}
\end{eqnarray}
The topological charge $Q$ is non trivial only in $D=1$ where:
\begin{equation}
Q={1\over 4\pi} \int\,dx\,dt\,{\bf \Omega}\cdot(\partial_x{\bf \Omega}
\times{\bf \dot\Omega})
\end{equation}
As a result, we obtain an effective NL$\sigma$M describing spin $S$ chains in (weak) 
uniform ($-$) or staggered ($+$) field. The microscopic derivation allows to obtain 
explicit expressions for the {\it bare} spin wave velocity $c=2Sa\sqrt{D}$ and stiffness 
$\rho_s=S^2a^{2-D}$ which coincide with those already known at ${\bf H}=0$.
While this derivation reproduces known results for a uniform field \cite{fisher}, 
it differs from the effective action
usually quoted in the literature, where the external field only couples to 
${\bf \Omega}$ via the Zeeman term. It is instructive to give a 
simple interpretation to
the formal result we have obtained: a staggered field ${\bf H}$ 
can be written as the sum of a
uniform field of the same strength minus a field twice as strong acting only on one
sublattice (say the sublattice $A$). In this way, by tracing out 
the degrees of freedom 
living on sublattice $B$ we obtain the same NL$\sigma$M 
action appropriate for a uniform field
with the addition of a Zeeman term $-2{\bf H}\cdot {\bf \Omega}$: 
This is exactly what we formally found in Eq. (\ref{acteffstag}).

In order to investigate the low energy spectrum of the NL$\sigma$M previously 
obtained (\ref{acteffstag}), we resort to a simple {\sl single mode approximation}:
the constraint ${\bf \Omega}^2=1$ may be lifted through the introduction
of a Lagrange multiplier $\lambda({\bf H})$ and a linear shift of the 
field ${\bf\Omega}$. In this way, the dynamical
correlation functions in imaginary time acquire a Lorenzian form:
\begin{eqnarray}
<S^z_{k,\omega}S^z_{-k,-\omega}>&=&\frac{cgS^2/a}{\omega^2+2cg\lambda+c^2k^2}
\label{smsa}\\
<S^{\pm}_{k,\omega} S^{\mp}_{-k,-\omega}>
&=&\frac{2cgS^2/a}{(\omega^2-{\bf H}^2)\mp 2i|{\bf H}|\omega+2cg\lambda+c^2k^2}
\nonumber
\end{eqnarray}
where $\lambda$ may be determined by a saddle point equation, 
or by fitting numerical data and $g=ca^{D-1}/\rho_s$. Here the $z$ axis identifies the
direction of the external field.
The poles of the correlation functions directly 
give the dispersion relation of the model.
Recalling that the  NL$\sigma$M describes the spin degrees of freedom on a single sublattice,
the wave vector $k$ is defined modulo $\pi$.
From expressions (\ref{smsa}) it is apparent that the correlation functions have single 
mode character with different dispersion relations 
in the transverse and longitudinal channel:
In particular, at each $k$ (modulo $\pi$) the transverse excitation splits into 
two different branches centered around the longitudinal dispersion:
\begin{eqnarray}
\Delta_L(k)&=&\sqrt{\Delta({\bf H})^2+c^2k^2}\nonumber\\
\Delta_T(k)&=&\Delta_L(k)\pm |\bf{H}|
\label{spectra}
\end{eqnarray}
This form of the energy spectrum should apply
to arbitrary spin $S$ at low energy and weak staggered fields.
It definitely differs from the predictions of the usual semiclassical treatments which
lead, within the same single mode approximation, 
to spectra of the form $\sqrt{\Delta({\bf H})^2+c^2k^2}$ in both channels. 
Interestingly, also the exact solution of the 1D $S=1/2$ $XY$ model in 
a staggered field along the
$z$ axis shows\cite{alcaraz} a similar excitation spectrum, providing some support to 
our semiclassical analysis. The isotropic $S=1/2$ Heisenberg model in a staggered field
has been also analyzed by bosonization techniques and conformal field theory 
\cite{oshikawa,alcaraz}.
The staggered field has been shown to open a gap both in the transverse and
in the longitudinal channel. This massive triplet is degenerate to leading order 
in the external field: $\Delta_T=\Delta_L\propto H^{2/3}$. It would be interesting 
to study the spliting of the gap by including subleading terms in order to compare
conformal field theory results with the NL$\sigma$M approach. 

In order to ascertain the validity of the single 
mode approximation in one dimensional
models, we performed Lanczos diagonalizations in $S=1$ chains. 
In particular we calculated the excitation spectrum by selecting, for each $k$, the 
excitation energy with the highest weight in the Lehmann representation of the
dynamical correlation function. 
\begin{figure}
\centerline{\psfig{bbllx=578pt,bblly=0pt,bburx=39pt,bbury=722pt,%
figure=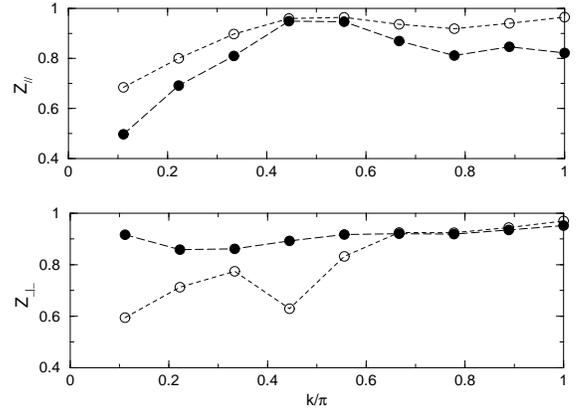,width=80mm,angle=-90}}
\caption{\baselineskip .185in \label{wei}
Longitudinal (above) and transverse (below) spectral weights for
$H=0.1$ (full dots) and $H=0.01$ (open dots) 
staggered field. Lanczos diagonalizations
in a N=18 lattice chain. In both cases, SMA is valid near $k=\pi$.}
\end{figure}

In Fig. \ref{wei} we show the largest normalized spectral weight $Z(k)$ 
as a function of the momentum $k$ of the excitation, 
showing that its value is always very close
to one for $k\sim \pi$: this implies that the sum rule ($\sum_n Z_n(k)=1$) is almost 
exhausted by a single excitation 
both in the transverse {\it and} in the longitudinal channel, thereby supporting
the single mode approximation usually adopted. 
Notice that the excitation with largest
weight does not always have the lowest energy, at fixed $k$.
Lanczos diagonalizations also show that 
the matrix element giving the spectral weight of the longitudinal 
excitation at $k=\pi$ decreases very quickly when the staggered field 
is switched on: It is lowered by a factor $2.5$ when the field reaches the
value $H=0.05$. This result should be compared with the modest 
decrease of the transverse spectral weight, which reduces just by $20\%$
in the same range. Both findings agree with the reported behavior of
neutron scattering data \cite{ray}.

\begin{figure}
\centerline{\psfig{bbllx=578pt,bblly=0pt,bburx=39pt,bbury=722pt,%
figure=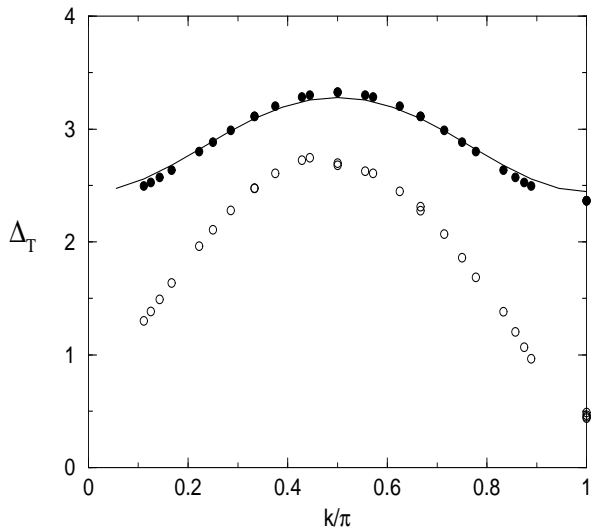,width=90mm,height=80mm,angle=-90}}
\caption{\baselineskip .185in \label{spe}
Transverse energy gap for two different field values: 
$H=0.01$ (open dots) and $H=1.0$ (full dots). 
Lanczos diagonalizations in N=12,14,16,18 chains.
In the high-field limit, the spectrum is well described by SWT (solid line).}
\end{figure}

In Fig. \ref{spe} the energy spectra are
shown for different field strength. While at low ${\bf H}$ the spectrum is
markedly asymmetrical around $k=\pi/2$, symmetry is restored at larger fields
where it closely approaches the form (\ref{swtd}) predicted by SWT. 
Data comes from Lanczos diagonalizations performed on chains with 
12, 14, 16 and 18 lattice sites.

\begin{figure}
\centerline{\psfig{bbllx=135pt,bblly=230pt,bburx=495pt,bbury=575pt,%
figure=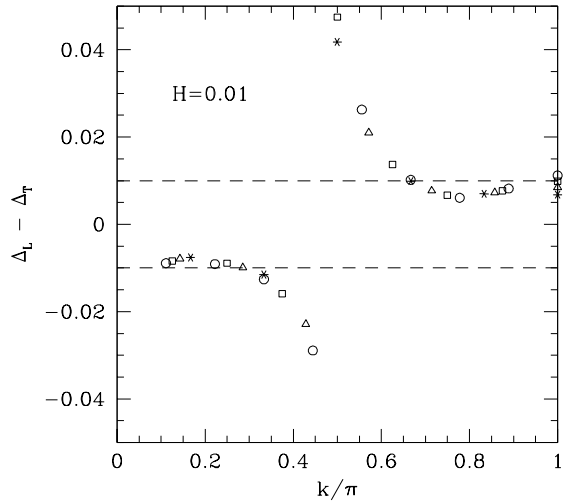,width=80mm,angle=0}}
\caption{\baselineskip .185in \label{gap}
Difference between the longitudinal and the transverse excitation
spectrum for $H=0.01$ and several chain  length (N=12,14,16,18).
The dashed lines correspond to the theoretical prediction 
$\Delta_L-\Delta_T=\pm H$.}
\end{figure}

A specific feature of our results is the
splitting of two branches in the transverse excitation spectrum when 
a weak staggered field is applied (\ref{spectra}). Fig. \ref{gap}
shows Lanczos diagonalizations data on several $S=1$ chains which
provide a numerical confirmation of the semiclassical predictions. 
For every wavevector $k$ we selected the longitudinal and the transverse
excitation with the largest spectral weight $Z$. When the continuum limit 
is appropriate, i.e. at sufficiently small $k$ (modulo $\pi$), 
two distinct excitation branches clearly appear, differing by $\pm |{\bf H}|$
from the longitudinal excitation, in agreement with the NL$\sigma$M analysis.
The noticeable deviations around $k=\pi/2$ are clearly due to lattice
effects which are not correctly reproduced in the continuum limit.
Notice that finite size corrections do not seem to affect the overall 
structure of the excitation spectrum of the model.

In Ref.\cite{yulu} the DMRG technique has been applied to investigate the lowest 
excitations in the longitudinal and transverse channels at $k=\pi$. 
The comparison of our results with DMRG is particularly instructive 
because it clearly shows the regimes where the use of the effective
action approach is justified. 

\begin{figure}
\centerline{\psfig{bbllx=50pt,bblly=26pt,bburx=573pt,bbury=730pt,%
figure=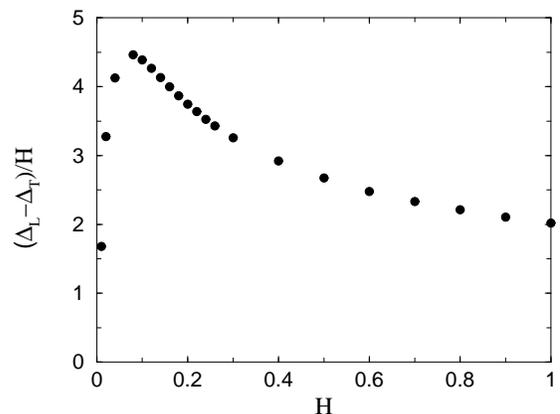,width=80mm,angle=-90}}
\caption{\baselineskip .185in \label{dmrg}
DMRG data\cite{yulu}: we plot the difference between the longitudinal gap and 
the lowest transverse excitation at $k=\pi$. 
The data are compatible with our theoretical predictions in both the 
low and strong field limits.}
\end{figure}

In fig. \ref{dmrg} we plot the quantity
$\gamma=(\Delta_L-\Delta_T)/H$ as a function of $ H $.
The numerical results identify two different behaviors: at strong fields
($H > 0.5$)  $\gamma$ saturates at $\gamma=2$, in agreement 
with SWT, while
in the low field limit ($H < 0.1$) the splitting between the longitudinal
and transverse gap quickly increases. In particular, the 
value $\gamma=1$ predicted by our semiclassical approach is compatible   
with the numerical data in the $H\to 0$ limit, although extended calculations 
in lower fields are required in order to validate our analysis.

\section{Spin-orbital models}
Spin orbital models have recently attracted considerable interest
in the attempt to explain the unusual magnetic 
properties of a class of quasi one dimensional materials, which 
includes C$_{60}$ compounds (eg. TDAE-C$_{60}$)\cite{auerbach}
and few metal oxides (eg. Na$_2$Ti$_2$SbO$_2$,
NaV$_2$O$_5$)\cite{pati}. The physical properties of these Mott
insulators are largely determined by the coupling between orbital
and spin degrees of freedom which may be dominated either by
Hund's rule or by dynamical Jahn-Teller effect. Possible 
realizations in higher dimensions are also found in fullerides
\cite{santoro} or in LiNiO$_2$ \cite{li}. The low energy physics
of these systems may be described by keeping only spin and orbital
degrees of freedom. If the orbital degeneracy is twofold, like in the
previous examples, the low energy model can be written in terms
of two sets of spin-$1/2$ operators per lattice site representing respectively
spin (S) and orbital (T) degrees of freedom.
Usually, spin isotropy in this effective low energy hamiltonian
is retained only in the physical spin variable S while 
terms which break rotational invariance in
the pseudo-spins T are generally allowed. However, the fully isotropic hamiltonian:
\begin{equation}
{\cal H} =J\,\sum_{<i,j>} \left [ {\bf S}_i\cdot {\bf S}_j + {\bf T}_i\cdot {\bf
T}_j\right ] + K \sum_{<i,j>} {\bf S}_i\cdot {\bf S}_j {\bf T}_i\cdot
{\bf T}_j
\label{st}
\end{equation}
has been the subject of several studies, particularly in
the two special cases $K=\pm 4J$. 
The $K=+4J$ hamiltonian can be written in terms
of permutators on each lattice site and may be relevant for
TDAE-C$_{60}$ while the $K=-4J$ 
naturally arises as the strong coupling limit of a microscopic
hamiltonian appropriate when dynamical Jahn-Teller effect prevails.
These two particular models have several remarkable properties: 
besides an obvious SU(2)$\times$SU(2) symmetry,
they are both invariant by a larger SU(4)
symmetry group \cite{santoro}. 
The 15 generators of the symmetry group include
the total spin and pseudospin operators: $\sum_R S^\alpha_R$
and $\sum_R T^\alpha_R$ and the further 9 operators
$\sum_R (\pm)^R S^\alpha_RT^\beta_R$ where the $\pm$ sign
corresponds to the two models $K=\pm 4J$. Note that, 
in the $-$ case, the SU(4) generators do not commute with
the translations by one lattice spacing although the 
hamiltonian does not break any symmetry of the lattice.
Remarkably, this model is also non frustrated: In a valence bond basis
the ground state can be shown to have 
positive semi-definite weights. This feature allows to 
perform very accurate Monte Carlo simulations on this system\cite{santoro2}.

Both models can be exactly solved in one 
dimensions \cite{martins,sutherland} with very different 
physical properties: the $K=+4J$ model ({\sl Sutherland model})
is gapless, with power law spin correlations whose leading
asymptotic behavior has a $x^{-3/2}$ decay and is characterized by 
oscillations with period equal to four lattice
spacing \cite{azaria}. Instead,
when $K=-4J$ ({\sl Valence Bond model}) the system spontaneously dimerizes,
the energy spectrum is gapped and correlations decay exponentially
\cite{martins,santoro}. Several relevant features of the
ground states of these hamiltonians have been argued to be 
applicable to wider regions in parameter space, also outside the
special SU(4) points \cite{itoi}. 

Here we will derive the effective low energy lagrangian
for both models in order to understand how such a different
physical behavior in one dimension may originate and to shed light 
on the phase diagram of the two models in higher dimensions.
In fact, analytical and numerical studies in two dimensions 
have suggested that SU(4) symmetry is not spontaneously broken 
in the ground state and a spin liquid phase may emerge\cite{santoro2,li}.
The general method developed in Section II can be straightforwardly
applied also to this class of hamiltonians 
defined on bipartite lattice because the interaction just couples
nearest neighbor sites. Let us discuss the two cases separately.

\subsection{Valence Bond model ($K=-4J$)}

The model has four orthogonal states per site in the lattice,
which correspond to the four possibilities $(\pm{1\over 2},\pm{1\over 2})$
for the $z$-components of the spin and pseudospin variables.
A set of coherent states is therefore labeled by a quartet of
complex numbers at each site $R$: $z_\alpha(R)$ 
obeying the normalization conditions $\sum_\alpha |z_\alpha(R)|^2=1$. 
We follow the convention to indicate the amplitude of the
$|\uparrow,\uparrow>$ state by $z_1$, of $|\uparrow,\downarrow>$ by
$z_2$, of $|\downarrow,\uparrow>$ by $z_3$ and of $|\downarrow,\downarrow>$
by $z_4$.  Using this representation, the partition function is written as 
\begin{eqnarray}
S_{eff}&=&\int_0^\beta dt \left [
\sum_{R}  \,z^*_\alpha(R,t)\dot z_\alpha(R,t) 
+<z(t)|H|z(t)> \right ] \nonumber \\
Z&=&\int {\cal D} \, z_\alpha(R,t) \,e^{-S_{eff}} 
\label{seff}
\end{eqnarray}
in close analogy with Eq. (\ref{action}). Summation over repeated 
labels $\alpha$ is understood. Again, by tracing out 
sublattice B, we reduce to an effective action defined only 
on sublattice A, formally given by:
\begin{equation}
S_{eff}=\int_0^\beta dt 
\sum_{R\in A}  \,z^*_\alpha(R,t)\dot z_\alpha(R,t)  
+\sum_{R\in B} F[{\bf K}(R,t)] 
\label{actionsu4}
\end{equation}
The functional $ F[{\bf K}(R,t)]$ is defined by
the single site problem in an external field:
\begin{eqnarray}
e^{-F[{\bf K}(R,t)]}={\rm Tr} \,U_\beta \nonumber\\
{dU_t\over dt} = - {\bf K}(R,t) \,U_t 
\label{onsitesu4}
\end{eqnarray}   
where ${\bf K}(R,t)$ is a $4\times 4 $ matrix whose components $K_{\mu\nu}$
depend on the classical field $z_\alpha(R^\prime,t)$
defined at nearest neighbor sites. The explicit form of
the matrix ${\bf K}(R,t)$ depends on the couplings of the hamiltonian, and
for the case we are examining is given by 
\begin{equation}
K_{\mu\nu}(R,t)=-J \sum_{\delta} \zeta_\mu^*(R+\delta,t)
\zeta_\nu (R+\delta,t)
\label{kkk}
\end{equation}
where, according to our conventions, the field $\zeta_\alpha$ is
simply related to the semiclassical variable $z_\alpha$ by:
$\zeta_1=z_4$, $\zeta_2=-z_3$, $\zeta_3=-z_2$ and $\zeta_4=z_1$.
To lowest order in the spatial derivatives, i.e. taking 
$z_\alpha(R,t)\sim z_\alpha(R+\delta,t)$, the instantaneous
ground state of ${\bf K}(R,t)$ is non degenerate and has components 
$\zeta^*_\alpha(R,t)$ with eigenvalue $\epsilon_0=-2DJ$
where $D$ is the spatial dimension and we specialized to 
hypercubic lattices. The three other eigenvalues of ${\bf K}(R,t)$
vanish and the corresponding eigenvectors
are any three four dimensional vectors orthogonal to the
ground state. Therefore the procedure outlined in Section II
can be straightforwardly applied and requires the evaluation of $\epsilon_0$
to second order in the lattice spacing $a$ (i.e. to second order 
in the spatial derivatives) which can be obtained by standard second order 
perturbation theory:
\begin{equation}
\epsilon_0=-2DJ + a^2 J \, {\bf \nabla} (z_\mu^* z_\nu) \cdot {\bf \nabla} (z_\mu z_\nu^*)
\label{eigst}
\end{equation}
Moreover the coefficient $\Gamma_{00}$ must be evaluated to
first order in $a$. The lowest order just cancels the Wess-Zumino
term, while the contribution linear in $a$ is non vanishing 
(for smooth configurations) only
in $D=1$, where it gives rise to a residual topological term:
\begin{equation}
\Gamma_{00}=z_\alpha \dot z^*_\alpha + a\,\partial_x (z_\alpha\dot z_\alpha^*)
+O(a^2)
\label{topost}
\end{equation}
Finally, the terms $\Gamma_{0j}(t)$ may be evaluated to lowest order in $a$.
For slowly varying fields $z_\alpha(t)$, only the $t^\prime\sim t$ region
does contribute to the integral in Eq. (\ref{free}) leading to:
\begin{equation} 
\int_0^t dt^\prime \sum_{j\ne 0} \Gamma_{0j}(t)\Gamma_{j0}(t^\prime)=
-{1\over 2DJ} \left \{ \dot z^*_\alpha \dot z_\alpha - 
z_\mu\dot z_\mu^* z^*_\nu\dot z_\nu \right \}
\label{f2st}
\end{equation}
Combining the results (\ref{eigst},\ref{topost},\ref{f2st}) we obtain the
required long wavelength limit of the effective semiclassical action
for the Valence Bond SU(4) model:
\begin{equation}
S_{eff} ={1\over 2g} \int dR \int dt \sum_{i=1}^{D+1}\sum_{\mu,\nu}
|\partial_i z_\mu^* z_\nu |^2  + i\pi Q
\label{cp3}
\end{equation}
The label $i$ runs over the $D+1$ space-time coordinates and
the continuum limit has been taken in the $D$ spatial directions while
the imaginary time variable has been suitably rescaled. 
The coupling constant $g$ is explicitly given by
\begin{equation}
g=a^{D-1}\sqrt{4D}
\label{gst}
\end{equation}
and the topological charge $Q$ is present only in $D=1$ where it reads:
\begin{equation}
Q = {1\over 2\pi i } \int dx\, dt\, \partial_x (z_\alpha\dot z^*_\alpha)
\label{topo}
\end{equation}

\begin{figure}
\centerline{\psfig{bbllx=135pt,bblly=230pt,bburx=495pt,bbury=575pt,%
figure=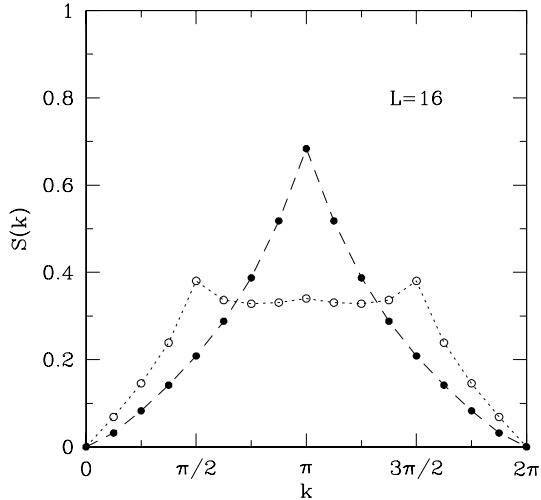,width=80mm,angle=0}}
\caption{\baselineskip .185in \label{sq}
Magnetic structure factor for the two SU(4) models considered here.
Lanczos data for a $16$-site chain with periodic boundary conditions.
Full dots: $K=-4J$ model. Open dots: $K=4J$ model.
}
\end{figure}

Tracing out a sublattice is an efficient way to take into account
the short range antiferromagnetic correlations present in the model
leading to an effective action describing the much smoother fluctuations
on a single sublattice. In order to support this interpretation of the
procedure we have adopted, Fig.(\ref{sq}) shows the magnetic structure factor 
obtained by Lanczos diagonalization on a $16$-site lattice: The sharp
peak at wavevector $k=\pi$ confirms that the most relevant correlations
have indeed periodicity of two lattice spacing.

The formal construction of the effective action for the Valence Bond
SU(4) model shows that the long wavelength and low energy physics
is described by a $D+1$ dimensional NL$\sigma$M or CP$^3$ model
(with topological 
angle $\theta=\pi$ in $D=1$). The known exact solution of the lattice model 
in one dimension implies that the CP$^3$ model at $g=2$ has correlations
exponentially decaying in space-time. Moreover, the spin model is known to 
develop dimer order in the thermodynamic limit, which implies breaking
of translational invariance by a lattice spacing together with breaking
of parity. In going to the continuum limit, one sublattice has been traced out
and then the CP$^3$ model remains translationally invariant but parity
breaking should still occur. This picture is supported by Lanczos
diagonalizations of the spin model showing that the quantum numbers of
the two SU(4) singlet states which collapse in the thermodynamic limit
correspond to momenta $P=0$ and $P=\pi$ and opposite parity (i.e.
reflections through a lattice site). Clearly, the mapping we have 
developed neglects cut-off effects and holds only for sufficiently
smooth configurations of the classical field. Therefore, the 
resulting estimate of the bare coupling constant $g=2$ should be taken
with caution but we are confident that the qualitative behavior of 
the CP$^3$ model does indeed capture the physics of this lattice spin
system, analogously to the familiar SU(2) case. 

The Valence Bond model we have considered 
belongs to the SU(n) class already studied by Affleck \cite{affleck} and 
Read and Sachdev \cite{read} by use of 1/n expansion. These analysis show
that the ground state breaks parity and translational symmetry 
and may be described by a Valence Bond Solid (VBS) in one 
dimension, at least for sufficiently large n. The exact solution
of the Valence Bond model \cite{martins} confirms that this picture 
holds down to n=4. The two dimensional 
case is more difficult: in the n$\to \infty$ limit the system has 
infinite degeneracy and can be represented as an arbitrary covering of the
lattice by nearest neighbor valence bonds. This degeneracy
is lifted at leading order in 1/n giving rise to
a (plaquette) resonating valence bond solid \cite{santoro}.
In fact, the model maps onto a dimer hamiltonian which
has been studied by Monte Carlo techniques \cite{runge}.
However, diagonalizations and Quantum Monte Carlo simulations
directly performed on the hamiltonian (\ref{st}) \cite{santoro2}
suggest that the ground state might be a spin liquid at n=4
and therefore argue in favor of a phase transition
between magnetically disordered phases as a function of n. 

\subsection{Sutherland model ($K=+4J$)}

The derivation of the NL$\sigma$M appropriate for the second
SU(4) point of the spin hamiltonian (\ref{st}) can be 
carried out following the same procedure adopted before 
and leading to Eqs. (\ref{seff}) and (\ref{actionsu4}).
However, the associated quantum mechanical problem 
(\ref{onsitesu4}) is now defined by the slightly different $4\times 4$ matrix
\begin{equation}
K_{\mu\nu}(R,t)=J \sum_{\delta} z_\mu(R+\delta,t)
z^*_\nu (R+\delta,t)
\label{kkk2}
\end{equation}
instead of Eq. (\ref{kkk}). To leading order in spatial fluctuations,
we can set $z_\mu(R+\delta)\sim z_\mu(R)$. In this case, the 
(instantaneous) ground state of the one site problem defined by the
matrix (\ref{kkk2}) is threefold degenerate. As a result, we cannot
carry on straightforwardly the trace over 
one sublattice, suggesting that this procedure is not able 
to eliminate the long range oscillations in spin correlations.
In turn, this means that the continuum limit cannot be taken
just by considering the spin configurations on a single sublattice, 
as in usual antiferromagnets and larger primitive cells must be
taken into account. A confirmation of such an interpretation comes from
Lanczos diagonalizations on this model: as shown in Fig. \ref{sq}, 
in one dimension the 
spin correlations displays oscillations characterized by the wavevector
$k=\pi/2$, implying a four site periodicity. This suggests a 
generalization of the procedure sketched in Section II: instead 
of tracing out one sublattice, we now keep one site every
four sites of the chain. Therefore we need the solution of
a three site problem, defined by the hamiltonian  (\ref{st}),
with ``time" dependent boundary conditions defined by the
classical field $z_\alpha(R,t)$ on the two adjacent sites. 
Equation (\ref{actionsu4}) is basically unchanged,
but now the sublattice $A$ includes only one fourth of the sites 
of the chain and the free energy functional $F$ is defined 
by Eq. (\ref{onsitesu4}) in terms of a 
matrix ${\bf K}$ acting on a Hilbert space of dimension $64$.
To lowest order in spatial fluctuations of the classical field 
$z_\alpha(R,t)$, the full spectrum of ${\bf K}$ can be explicitly
obtained by solving the eigenvalue equation:
\begin{equation}
J\left [z_\mu z^*_\alpha\ u_{\alpha,\nu,\lambda} +u_{\nu,\mu,\lambda}
+u_{\mu,\lambda,\nu}+z_\lambda z^*_\alpha u_{\mu,\nu,\alpha}\right ]
=\epsilon\,u_{\mu,\nu,\lambda}
\label{3site}
\end{equation}
In particular, the ground state wavefunction of the quantum problem 
is now non-degenerate and reads 
\begin{equation}
u_{\mu,\nu,\lambda}={1\over \sqrt{6}} z^*_\alpha\epsilon_{\alpha\mu\nu\lambda}
\label{3gs}
\end{equation}
where $\epsilon_{\alpha\mu\nu\lambda}$ is the fully antisymmetric
Levi-Civita tensor. The corresponding eigenvalue is simply $\epsilon_0=-2J$,
while the Berry phase contribution gives 
\begin{equation}
\Gamma_{00}={1\over 6} z_\alpha 
\epsilon_{\alpha\mu\nu\lambda} 
\dot z^*_\beta \epsilon_{\beta\mu\nu\lambda}= 
z_\alpha\dot z^*_\alpha
\end{equation}
and cancels the Wess-Zumino term in the effective action.
In order to obtain a non-trivial theory we have to include 
long wavelength fluctuations of the classical field $z(R,t)$.
This can be done by use of perturbation theory in the 
associated quantum three site problem. The solution is
sketched in Appendix where the required terms are 
explicitly evaluated. Here we just quote the final
result, after having taken the continuum limit:
The spin fluctuations on the sublattice are described 
by a NL$\sigma$M which coincides with the one obtained 
for the Valence Bond model (\ref{cp3}) at a bare coupling
\begin{equation}
g={6\over \sqrt{5}}
\label{gc}
\end{equation}
larger than the estimate $g=2$ obtained in one dimension 
for the Valence Bond model (\ref{gst}). Also for this
hamiltonian the topological angle $\theta$ is given by
$\theta=\pi$.

As previously stressed, the Sutherland model is critical 
in one dimension, with power law correlations characterized
by oscillations of period $4a$ corresponding to a characteristic wavevector
$k=\pi/2$, as also confirmed by the Lanczos diagonalization results 
shown in Fig. \ref{sq}. If the mapping onto a CP$^3$
theory (with topological angle $\theta=\pi$) faithfully
describes the long wavelength physics of the lattice model,
we conclude that the CP$^3$ action in $1+1$ dimension has
correlations 
\begin{equation}
<z^*_\alpha(0,0) z_\beta(0,0) 
z^*_\beta(x,t)z_\alpha(x,t)> \sim (x^2+t^2)^{-3/4}
\end{equation}
On the other hand, the exact solution of the Valence Bond hamiltonian
together with the semiclassical mapping of Section IV-A
implies that the CP$^3$ model has exponentially decaying correlations
at $g=2$. These two results may be reconciled if we assume that the
CP$^3$ theory in $1+1$ dimension undergoes a second order 
phase transition as a function of the coupling $g$
and the Sutherland model describes the physics at 
the critical point. Another possibility would be the
occurrence of a gapless {\it phase} in the model, like
in the celebrated CP$^1$ case. This alternative explanation,
however, conflicts with available analytical
and numerical evidence pointing towards a massive regime
at strong coupling \cite{cp}. 
A phase transition in the CP$^3$ model
cannot be related to a spontaneous breaking of the
continuous SU(4) symmetry while a possibility is
the occurrence of parity breaking in one of the 
two phases. Of course the estimate (\ref{gc}) we have obtained for
the critical coupling will be renormalized by finite cut-off
effects but the overall picture of the phase diagram 
emerging from the semiclassical analysis should be robust.
This result may have implications in the framework of 
the strong CP problem in field theory. 
\cite{cp}.

\section{Conclusions}
In this paper we discussed some application of the semiclassical approach
to spin systems in low dimensionality. This technique is known to 
capture the qualitative features of quantum models and to 
provide a useful framework for the interpretation of  experimental
and numerical data. 

For Heisenberg chains in a staggered field
we pointed out some difficulty and  inconsistency of usual treatments,
especially in the weak field limit, where the structure of the effective
NL$\sigma$M turns out to be richer than expected. The general form
of the excitation spectrum predicted by semiclassical approaches
has been confirmed by use of Lanczos diagonalizations in finite clusters
which also provide some support to the usually adopted SMA.
Our results are fully compatible with existing DMRG data and
show that different physical regimes occur in the 
phase diagram of this model. A recent DMRG investigation of a
$S=2$ spin chain also pointed out a similar behavior \cite{capone}.
The new NL$\sigma$M we 
derived is expected to represent only the {\sl low field} region
while, for moderate/high staggered fields, simple perturbative
approaches, like spin wave theory, are fully adequate to describe 
the excitation spectrum of the model. It seems unlikely that a 
single effective action in the continuum limit might be able to
encompass these two very different physical behaviors.

A semiclassical analysis of
spin-orbital models characterized by a SU(4) symmetry
has also been performed in two different regimes.
In one case we straightforwardly obtained, in the long wavelength limit, 
a mapping to the CP$^3$ NL$\sigma$M describing the fluctuations of
spin-orbitals degrees of freedom on the same sublattice. Recent numerical
simulations argued in favor of a disordered ground state in such a model
for $D=2$, while in one dimension the exact solution of the lattice hamiltonian
proved that spontaneous dimerization occurs. In the other SU(4) model
we have examined, a smooth continuum limit requires a coarse graining
over several lattice sites, suggesting that the relevant fluctuations
are characterized by a wavevector $k$ different from the antiferromagnetic one.
We explicitly performed the analysis only for the one dimensional model, where 
we again obtained the same CP$^3$ model which now describes fluctuations
about $k=\pi/2$. The phase diagram of such a  NL$\sigma$M has been extensively
studied, particularly in $1+1$ dimension, in connection to the CP problem in field
theory. Several proposals
have been put forward in the literature, including spontaneous parity 
breaking and deconfining transition.
The long wavelength mapping between spin chains and two dimensional
field theories may provide a clue for the final understanding
of the phase diagram of the CP$^n$ model, analogously to what has been
found for the Wess Zumino Witten model.

We gratefully acknowledge useful correspondence with A.Pelissetto
and G. Morandi. We also thank Yu Lu and Jizhong Lou for providing DMRG data. 
\section{Appendix}
In this appendix we briefly illustrate the procedure adopted for the 
analytical solution of the generalized three site problem 
\begin{equation}
J\left [z_\mu z^*_\alpha\ u_{\alpha,\nu,\lambda} +u_{\nu,\mu,\lambda}
+u_{\mu,\lambda,\nu}+\bar z_\lambda \bar z^*_\alpha u_{\mu,\nu,\alpha}\right ]
=\epsilon\,u_{\mu,\nu,\lambda}
\end{equation}
up to second order in the lattice spacing $a$. Here $\bar z_\alpha=
z_\alpha + 4a \partial_x z_\alpha +8a^2\partial_x^2 z_\alpha + O(a^3)$.
We employ standard second order perturbation theory which gives
\begin{equation}
\Delta\epsilon=<u_0|\Delta {\bf K} |u_0> + \sum_{n\ne 0} {
|<u_0|\Delta {\bf K} |u_n>|^2\over \epsilon_0-\epsilon_n}
\label{pert}
\end{equation}
The unperturbed eigenvectors $|u_n>$ are the solutions of the
eigenvalue equation (\ref{3site}) and $\epsilon_n$ are the
corresponding eigenvalues. The ground state $|u_0>$ is explicitly given
in Eq.(\ref{3gs}) and $\epsilon_0=-2J$. Due to the manifest SU(4) 
invariance of the eigenvalue equation, the external field $z_\alpha$ may be
chosen to point in the ``1" direction 
without loss of generality: $z_\alpha=\delta_{\alpha,1}$. 
The lowest order term gives:
\begin{equation}
<u_0|\Delta {\bf K} |u_0>=
{16\over 3}a^2J \left [ \,\vert (z^*_\alpha\partial_x z_\alpha)
\vert^2 - \partial_x z_\alpha \partial_x z^*_\alpha\,\right ]
\end{equation}
while the sum over excited states which appears at second order
requires the evaluation of the matrix element
\begin{equation}
<u_0|\Delta {\bf K} |u_n>= -{4aJ\over \sqrt{6}} \epsilon_{1\alpha\mu\nu} 
u_{\mu\nu 1} \partial_x z_\alpha
\end{equation}
for a generic excited state $u_{\mu\nu 1}$. The only solutions of the 
unperturbed eigenvalue equation (\ref{3site}) which give a non vanishing
contribution are those corresponding to the eigenvalue $\epsilon=\pm\sqrt{2}\,J$
and to $\epsilon=0$. The former states are given by $u_{123}=
u_{231}=-u_{132}=-u_{321}=\pm u_{312}/\sqrt{2}=\mp u_{213}/\sqrt{2}=1/\sqrt{8}$.
The latter states are $u_{123}=u_{321}=-u_{132}=-u_{231}=1/2$
Both excited states are three times degenerate: the other states  
being obtained by cyclic permutations of the labels $(234)$.
Inserting these results into the perturbative expansion (\ref{pert})
we get:
\begin{equation}
\Delta\epsilon=-{4\over 3} J\,a^2 \left [ \,\vert (z^*_\alpha\partial_x z_\alpha)
\vert^2 - \partial_x z_\alpha \partial_x z^*_\alpha\,\right ]
\end{equation}
The last required step is the evaluation of the Berry phase term 
$\Gamma_{00}$ to first order in the lattice spacing. 
Again, by use of perturbation theory, we get
\begin{equation}
\Gamma_{00}=<u_0|\dot u_0>+2i\,{\rm Im} \sum_{n\ne 0} {
<u_0|\Delta {\bf K} |u_n>\over \epsilon_0-\epsilon_n}
<u_n|\dot u_0>
\end{equation}
The intermediate states which contribute to the sum are those
corresponding to $\epsilon=\pm\sqrt{2}\, J$ which give:
\begin{equation}
\Gamma_{00}=z_\alpha\dot z^*_\alpha -2a\,\partial_x(\dot z^*_\alpha z_\alpha)
\end{equation}
This, in turn, gives the well known topological term 
quoted in the text (\ref{topo}).

\end{document}